\documentclass[pra,aps,showpacs,onecolumn,twoside,superscriptaddress]{revtex4}

\usepackage{amsmath,amsfonts,amssymb,color,epsfig,graphics,graphicx,latexsym,revsymb,theorem,url,verbatim}

\newtheorem{definition}{Definition}
\newtheorem{proposition}[definition]{Proposition}
\newtheorem{lemma}[definition]{Lemma}

\newtheorem{theorem}[definition]{Theorem}
\newtheorem{corollary}[definition]{Corollary}
\newtheorem{conjecture}[definition]{Conjecture}

\newtheorem{remark}[definition]{Remark}
\newtheorem{example}[definition]{Example}

\def\squareforqed{\hbox{\rlap{$\sqcap$}$\sqcup$}}
\def\qed{\ifmmode\squareforqed\else{\unskip\nobreak\hfil
\penalty50\hskip1em\null\nobreak\hfil\squareforqed
\parfillskip=0pt\finalhyphendemerits=0\endgraf}\fi}
\def\endenv{\ifmmode\;\else{\unskip\nobreak\hfil
\penalty50\hskip1em\null\nobreak\hfil\;
\parfillskip=0pt\finalhyphendemerits=0\endgraf}\fi}
\newenvironment{proof}{\noindent \textbf{{Proof.~} }}{\qed}
\def\Dbar{\leavevmode\lower.6ex\hbox to 0pt
{\hskip-.23ex\accent"16\hss}D}
\makeatletter
\def\url@leostyle{%
  \@ifundefined{selectfont}{\def\UrlFont{\sf}}{\def\UrlFont{\small\ttfamily}}}
\makeatother
\def\bcj{\begin{conjecture}}
\def\ecj{\end{conjecture}}
\def\bcr{\begin{corollary}}
\def\ecr{\end{corollary}}
\def\bd{\begin{definition}}
\def\ed{\end{definition}}
\def\bea{\begin{eqnarray}}
\def\eea{\end{eqnarray}}
\def\bem{\begin{enumerate}}
\def\eem{\end{enumerate}}
\def\bex{\begin{example}}
\def\eex{\end{example}}
\def\bim{\begin{itemize}}
\def\eim{\end{itemize}}
\def\bl{\begin{lemma}}
\def\el{\end{lemma}}
\def\bpf{\begin{proof}}
\def\epf{\end{proof}}
\def\bpp{\begin{proposition}}
\def\epp{\end{proposition}}
\def\br{\begin{remark}}
\def\er{\end{remark}}
\def\bt{\begin{theorem}}
\def\et{\end{theorem}}

\newcommand{\nc}{\newcommand}

\def\a{\alpha}
\def\b{\beta}

\def\d{\delta}
\def\e{\epsilon}

\def\z{\zeta}

\def\l{\lambda}

\def\r{\rho}
\def\s{\sigma}

\def\ph{\varphi}

\def\ps{\psi}

\def\G{\Gamma}

\nc{\bbC}{{\mathbb{C}}}

\nc{\cA}{{\cal A}} \nc{\cB}{{\cal B}} \nc{\cC}{{\cal C}}
\nc{\cD}{{\cal D}} \nc{\cE}{{\cal E}} \nc{\cF}{{\cal F}}
\nc{\cG}{{\cal G}} \nc{\cH}{{\cal H}} \nc{\cI}{{\cal I}}
\nc{\cJ}{{\cal J}} \nc{\cK}{{\cal K}} \nc{\cL}{{\cal L}}
\nc{\cM}{{\cal M}} \nc{\cN}{{\cal N}} \nc{\cO}{{\cal O}}
\nc{\cP}{{\cal P}} \nc{\cR}{{\cal R}} \nc{\cS}{{\cal S}}
\nc{\cT}{{\cal T}} \nc{\cU}{{\cal U}} \nc{\cV}{{\cal V}}
\nc{\cW}{{\cal W}} \nc{\cX}{{\cal X}} \nc{\cZ}{{\cal Z}}

\nc{\hA}{{\hat{A}}} \nc{\hB}{{\hat{B}}} \nc{\hC}{{\hat{C}}}
\nc{\hD}{{\hat{D}}} \nc{\hE}{{\hat{E}}} \nc{\hF}{{\hat{F}}}
\nc{\hG}{{\hat{G}}} \nc{\hH}{{\hat{H}}} \nc{\hI}{{\hat{I}}}
\nc{\hJ}{{\hat{J}}} \nc{\hK}{{\hat{K}}} \nc{\hL}{{\hat{L}}}
\nc{\hM}{{\hat{M}}} \nc{\hN}{{\hat{N}}} \nc{\hO}{{\hat{O}}}
\nc{\hP}{{\hat{P}}} \nc{\hR}{{\hat{R}}} \nc{\hS}{{\hat{S}}}
\nc{\hT}{{\hat{T}}} \nc{\hU}{{\hat{U}}} \nc{\hV}{{\hat{V}}}
\nc{\hW}{{\hat{W}}} \nc{\hX}{{\hat{X}}} \nc{\hZ}{{\hat{Z}}}

\def\dim{\mathop{\rm Dim}}

\def\max{\mathop{\rm max}}
\def\min{\mathop{\rm min}}

\def\rank{\mathop{\rm rank}}

\def\tr{\mathop{\rm Tr}}

\def\GL{{\mbox{\rm GL}}}

\def\bigox{\bigotimes}
\def\dg{\dagger}

\def\op{\oplus}
\def\ox{\otimes}
\def\ra{\rightarrow}

\def\sue{\subseteq}

\newcommand{\bra}[1]{\langle#1|}
\newcommand{\ket}[1]{|#1\rangle}
\newcommand{\proj}[1]{| #1\rangle\!\langle #1 |}
\newcommand{\ketbra}[2]{|#1\rangle\!\langle#2|}
\newcommand{\braket}[2]{\langle#1|#2\rangle}

\newcommand{\jmp}{J. Math. Phys.}
\newcommand{\jpa}{J. Phys. A}

\newcommand{\pla}{Phys. Lett. A}

\def\bR{{\mbox{\bf R}}}

\def\bC{{\mbox{\bf C}}}

\begin{document}
\title{Qubit-qudit states with positive partial transpose}

\author{Lin Chen}
\affiliation{Department of Pure Mathematics and Institute for
Quantum Computing, University of Waterloo, Waterloo, Ontario, N2L
3G1, Canada} \affiliation{Centre for Quantum Technologies, National
University of Singapore, 3 Science Drive 2, Singapore 117542}
\email{cqtcl@nus.edu.sg (Corresponding~Author)}

\def\Dbar{\leavevmode\lower.6ex\hbox to 0pt
{\hskip-.23ex\accent"16\hss}D}
\author {{ Dragomir {\v{Z} \Dbar}okovi{\'c}}}

\affiliation{Department of Pure Mathematics and Institute for
Quantum Computing, University of Waterloo, Waterloo, Ontario, N2L
3G1, Canada} \email{djokovic@uwaterloo.ca}

\begin{abstract}
We show that the length of a qubit-qutrit separable state is
equal to $\max(r,s)$, where $r$ is the rank of the state and
$s$ the rank of its partial transpose. We refer to the
ordered pair $(r,s)$ as the birank of this state. We also
construct examples of qubit-qutrit separable states of any
feasible birank $(r,s)$. We determine the closure of the set
of normalized two-qutrit entangled states having positive
partial transpose (PPT) of rank four. The boundary of this
set consists of all separable states of length at most four.
We prove that the length of any qubit-qudit separable state
of birank $(d+1,d+1)$ is equal to $d+1$. We also show that
all qubit-qudit PPT entangled states of birank $(d+1,d+1)$
can be built in a simple way from edge states. If $V$ is a
subspace of dimension $k<d$ in a $2\ox d$ space such that
$V$ contains no product vectors, we show that the set of
all product vectors in $V^\perp$ is a vector bundle of rank
$d-k$ over the projective line. Finally, we explicitly
construct examples of qubit-qudit PPT states (both separable
and entangled) of any feasible birank.
\end{abstract}

\date{ \today }

\pacs{03.67.Mn, 03.65.Ud}



\maketitle



%
%
%
%
%
%

\section{\label{sec:introduction} Introduction}

Bipartite quantum states are key ingredients in many fundamental
applications and theoretical problems of quantum information. Bell
states are pure bipartite states and useful for teleportation
\cite{bbc93} and dense coding \cite{bw92}. It has been shown by
experiment \cite{bell64,agr82} that Bell states violate the Bell
inequality. So it indicates the nonlocality, which is an essential
feature of quantum physics. Unfortunately, there is no pure state
existing in nature, as it extremely quickly turns into a mixed state
due to the decoherence from the environment. Extraction of Bell
states, as original quantum resource, from mixed states under local
operations and classical communication (LOCC) is known as
entanglement distillation. It is a central task in entanglement
theory \cite{bds96}. This task is also the key method for
constructing the distillable key, which supports the security proof
in quantum cryptography \cite{sp00}. Entanglement distillation is
possible only if the mixed state is entangled. A non-entangled
state, also known as a separable state, is by definition a convex
sum of product states \cite{werner89}. Such states can be prepared
locally in experiments. It is natural to pose the separability
problem, i.e., to ask whether a given state is separable. It is
known in computational complexity theory \cite{gurvits03} that this
problem is NP-hard. Actually, both the entanglement distillation and
separability problem cannot be effectively solved even for bipartite
states (for recent progress in a particular case see
\cite{cd11JPA}).

For a bipartite state $\r$ acting on the Hilbert space
$\cH:=\cH_A\ox \cH_B$, the partial transpose computed in an orthonormal (o.n.) basis $\{\ket{a_i}\}$ of system A, is defined
by $\r^\G=\sum_{ij}\ketbra{a_j}{a_i}\ox\bra{a_i}\r\ket{a_j}$. The dimensions of $\cH_A$ and $\cH_B$ are denoted by $M$ and
$N$, respectively. We say that $\r$ is a $k\times l$ {\em state}
if its local ranks are $k$ and $l$, i.e., $\rank\r_A=k$ and
$\rank\r_B=l$.
We say that $\r$ is a PPT [NPT] state if $\r^\G\ge0$ [$\r^\G$ has at least one negative eigenvalue]. Evidently, a separable
state must be PPT. The converse is true only if $MN\le6$
\cite{peres96,hhh96}, in which case the separability problem is
solved. The first examples of two-qutrit PPT entangled states (PPTES) were constructed in purely mathematical context by Choi
and St{\o}rmer in the 1980s \cite{choi80,stormer82}. They were introduced into quantum information theory in 1997 \cite{horodecki97}. The full description of two-qutrit PPTES of rank four was constructed in 2011 in \cite{cd11JMP} and \cite{skowronek11} (independently). The most intriguing feature of
PPTES is that they are not distillable, i.e., they cannot be converted into Bell states under LOCC. So PPTES are not directly useful for entanglement distillation. Nevertheless, some PPTES
can be used to construct distillable key \cite{hhh05}.

In the bipartite setting, $2\times N$ states are related to many
problems in quantum information and have received a lot of
attention.

First, one of the most known analytical formulas for entanglement
measures is the entanglement of formation of two-qubit states
\cite{wootters98}. Part of the derivation of this formula relies on
the observation that the two-qubit separable states have
\textit{length} at most four. The length of a separable state $\r$,
denoted by $L(\r)$, is defined as the minimal number of pure product
states whose mixture is $\r$ \cite{dtt00}. So it represents the
minimal physical efforts that realize $\r$ by the entanglement of
formation. Two separable states with different length are not
equivalent under stochastic local operations and classical
communications (SLOCC) \cite{dvc00}.

On the other hand, the purification of a $2\times N$ separable state
$\r$ of rank $r$ is a $2\times N \times r$ tripartite pure state
$\ket{\ps}$. So the tensor rank of the latter is not larger than the
length of $\r$ \cite{ccd10}. This connection is computationally
operational since the tensor rank of $\ket{\ps}$ can be computed by
efficient programs \cite{jaja78,cms10}.

Second, a first systematic study of $2\times N$ PPT states $\r$
was published in 1999 \cite{kck00}. Their main result is that
$\r$ is separable when its rank is equal to $N$. Recently,
$2\times4$ extremal PPTES for various biranks have been
constructed in \cite{agk10}. Such states are extreme points
of the set of PPT states, and have been studied in bipartite
systems of arbitrary dimensions \cite{cd12}. Entanglement
witnesses for physically detecting entanglement of $\r$ have
been also studied \cite{atl11}.

Third, all $2\times N$ NPT states are distillable \cite{dss00}, while the distillability of $3\times 3$ NPT states still remains as a major open problem in entanglement theory.

Fourth, it has been shown that the $2\times N$ states contain
quantum correlation measured by quantum discord \cite{bc10}.

Motivated by a desire for deeper understanding of these results and
their possible applications to various quantum-information tasks and
to computational complexity, we continue in this paper the
investigation of $2\times N$ separable states and PPTES. After a
preliminary technical Lemma \ref{le:connected}, we prove in
Corollary \ref{cor:birank(r,r)} that given a $2\times N$ separable
state $\s$ we can subtract from it a pure product state to obtain
another PPT state of lower birank. This result is essential for the
computation of the length of a $2\times3$ separable state $\r$ of
given birank $(r,s)$. Namely, we show in Proposition
\ref{prop:SEP2x3} that $L(\r)=\max\{r,s\}$. We give in Table
\ref{tab:SEP2x3} concrete examples of separable states $\r$ for all
possible lengths and biranks. Similar results for two-qubit
separable states are shown in Table \ref{tab:SEP2x2}. By using these
result and new  Lemmas \ref{le:sep4}, \ref{le:SEPrank3} and
\ref{le:SEP(4,4)}, we determine the closure of the set $\cE$ of
normalized two-qutrit PPTES of rank four (see Theorem
\ref{thm:Closure}). It turns out that this closure is the union of
$\cE$ and the set $\cS'_4$ of separable states of length at most
four. In Example \ref{ex:3x3pptes}, we construct a two-qutrit
separable state $\r$ of rank five, such that whenever
$\s=\r-\proj{e,f}$ is a PPT state of birank equal to $(r-1,s)$,
$(r,s-1)$ or $(r-1,s-1)$, then $\s$ is necessarily entangled. This
fact can be regarded in physics as the loss of separability by
subtraction of a pure product state. In Theorem
\ref{thm:SEP2xN,(N+1,N+1)}, we show that the $2\times N$ separable
state of birank $(N+1,N+1)$ has length $N+1$. In the same theorem we
show that a $2\times N$ PPTES $\r$ of birank $(N+1,N+1)$ must be the
B-direct sum of several pure product states and an edge state $\s$
\cite{sbl01}. So two $2\times N$ PPTES $\r_1$ and $\r_2$ of birank
$(N+1,N+1)$ are equivalent under SLOCC only if the edge states
$\s_1$ and $\s_2$, and the pure product states are simultaneously
equivalent under SLOCC. This is a new method to the hard problem of
deciding equivalent mixed states. Furthermore, the entanglement
witness detecting the entanglement of the edge state $\s$ would be
able to detect the entanglement of the PPTES $\r$.

In Proposition \ref{pp:Prva} we study the set of all product vectors
contained in the orthogonal complement $V^\perp$ of a completely
entangled space $V$ of dimension $k<N$. We show that this set is a
vector bundle of rank $N-k$ over the projective line. In the special
case $k=N-1$, its projectivization is a rational normal curve, a
well known object in classical algebraic geometry. In Propositions
\ref{pp:2xN,PPTbirank} and \ref{pp:PPTES,(N+1+p,N+1+k)}, we prove
the existence of $2\times N$ separable as well as PPT entangled
states having birank $(r,s)$, where $r$ and $s$ are arbitrary
inetegers in the range $N+1,\ldots,2N$. The proofs are based on
Proposition \ref{pp:Prva} and the recently constructed PPTES in
\cite{tah12}. Finally in Example \ref{ex:negativeEIGENVALUE}, for
each $m\in\{1,\ldots,N-1\}$, we construct a $2\times N$ NPT state
whose partial transpose has exactly $m$ negative eigenvalues.

The paper is organized as follows. In Sec. \ref{sec:preliminary}
we state the known facts which we often use in this paper. In
Sec. \ref{sec:Lengths2x3} we solve the length problem for
$2\times3$ separable states. The main result is presented in
Proposition \ref{prop:SEP2x3}. In Sec. \ref{sec:equivalence} we determine the closure of $3\times3$ PPTES of rank four. The main
result is stated in Theorem \ref{thm:Closure}. In Sec. \ref{sec:existence} we study the $2\times N$ PPT states of prescribed rank. The main results are presented in Theorem \ref{thm:SEP2xN,(N+1,N+1)}, Proposition \ref{pp:Prva}, Proposition
\ref{pp:2xN,PPTbirank} and \ref{pp:PPTES,(N+1+p,N+1+k)}.

\section{\label{sec:preliminary} Preliminaries}

We shall write $I_k$ for the identity $k\times k$ matrix. We denote
by $\cR(\r)$ and $\ker \r$ the range and kernel of a linear map
$\r$, respectively. From now on, unless stated otherwise, the states
will not be normalized. We shall denote by
$\{\ket{i}_A:i=0,\ldots,M-1\}$ and $\{\ket{j}_B:j=0,\ldots,N-1\}$
o.n. bases of $\cH_A$ and $\cH_B$, respectively. The subscripts A
and B will be often omitted. 
%
For any bipartite state $\r$ we have
 \bea
 \label{ea:rhoBGamma=rhoB}
\left( \r^\G \right)_B &=& {\tr}_A \left( \r^\G \right) =
{\tr}_A \r = \r_B, \\
 \label{ea:rhoAGamma=rhoGammaA}
\left( \r^\G \right)_A &=& {\tr}_B \left( \r^\G \right) = \left(
{\tr}_B \r \right)^T = ( \r_A )^T. \eea Here the exponent T denotes
transposition. Consequently,
 \bea
\rank \left( \r^\G \right)_{A,B} = \rank \r_{A,B}.
 \eea
If $\r$ is an $M\times N$ PPT state, then $\r^\G$ is too. If $\r$ is
a PPTES so is $\r^\G$, but they may have different ranks. An example
is the two-qubit separable state of birank $(3,4)$, see Table
\ref{tab:SEP2x2}.

Let us now recall some basic results from quantum information
regarding the separability and PPT properties of bipartite states.
Let us start with the basic definition.

 \bd
We say that two $n$-partite states $\r$ and $\s$ are {\em equivalent
under stochastic local operations and classical communications}
({\em SLOCC-equivalent} or just {\em equivalent}) if there exists an
invertible local operator (ILO) $A=\bigox^n_{i=1} A_i \in
\GL:=\GL_{d_1}(\bC)\times\cdots\times\GL_{d_n}(\bC)$ such that
$\r=A\s A^\dg$ \cite{dvc00}.
 \ed

In most cases of the present work, we will have $n=2$. It is easy
to see that any ILO transforms PPT, entangled, or separable state
into the same kind of states. The length of a separable state is
invariant under ILO and is non-increasing under all local operations. We shall often use ILOs to simplify the density
matrices of states. We say that a subspace of $\cH$ is {\em
completely entangled} (CES) if it contains no product vectors. We
require product vectors to be nonzero. For counting purposes we
do not distinguish product vectors which are scalar  multiples of
each other.

We recall that $D=d_1d_2\cdots d_n-\sum^n_{i=1} d_i+n-1$ is the
maximal dimension of CES in $d_1\ox\cdots\ox d_n$ \cite{parthasarathy04}. It follows easily from \cite[Theorem 60]{cd12}
that any CES is contained in one of dimension $D$.

The first assertion of the following theorem is \cite[Theorem
23]{cd11JPA}. The second one follows from its proof where the
parameter $a$ was only shown to be real and nonzero. The stronger
claim that (like $b,c,d$) $a$ can also be chosen to be positive has
been proved in \cite[Theorem 7]{cd12JMP}.

 \bpp
 \label{prop:3x3rank4PPTES,invariantexpression}
 $(M=N=3)$ Any $3\times3$ PPTES $\r$ of rank four is SLOCC-equivalent
to one which is invariant under partial transpose, i.e., there exist
$A,B\in\GL_3(\bC)$ such that $\s:=A \ox B~\r~A^\dg \ox B^\dg$
satisfies the equality $\s^\G=\s$. Moreover, we may assume that
$\s=C^\dag C$ where $C=[C_0~C_1~C_2]$ and
 \bea \label{ea:Blokovi}
 C_0= \left[\begin{array}{ccc}
0 & a & b \\
0 & 0 & 1 \\
0 & 0 & 0 \\
0 & 0 & 0
\end{array}\right],\quad
 C_1= \left[\begin{array}{ccc}
0 & 0 & 0 \\
0 & 0 & c \\
0 & 0 & 1 \\
1 & 0 & -1/d
\end{array}\right],\quad
 C_2= \left[\begin{array}{ccc}
0 & -1/b & 0  \\
0 & 1 & 0  \\
1 & -c & 0 \\
d & 0 & 0
\end{array}\right]; \quad a,b,c,d>0.
\eea
 \epp
This equation will be used to show that separable states of length
at most four are in the closure of the set of non-normalized
$3\times3$ PPTES of rank four in Lemma \ref{le:sep4}. To prove this
lemma we will need the definition of the term ``general position''
\cite[Definition 7]{cd12}.

 \bd \label{def:GenPos} We say that a family of product vectors
$\{\ket{\psi_i}=\ket{\phi_i}\ox\ket{\chi_i}:i\in I\}$ is in {\em
general position} (in $\cH$) if for any $J\sue I$ with $|J|\le M$
the vectors $\ket{\phi_j}$, $j\in J$, are linearly independent and
for any $K\sue I$ with $|K|\le N$ the vectors $\ket{\chi_k}$, $k\in
K$, are linearly independent.
 \ed

The next result is from \cite[Theorem 1]{hst99}. It is useful in the
characterization of the length of $2\times3$ separable states.
 \bt
\label{thm:PPTMxNrank<M,N} The $M\times N$ states of rank less than
$M$ or $N$ are 1-distillable, and consequently they are NPT.
 \et

The next result follows from \cite[Theorem 10]{cd11JPA},
\cite{hlv00} and Theorem \ref{thm:PPTMxNrank<M,N}, see also
\cite[Proposition 6 (ii)]{cd11JPA}.

 \bpp
 \label{prop:PPTMxNrankN}
Let $\r$ be an $M\times N$ state of rank $N$.

(i) If $\r$ is PPT, then it is a sum of $N$ pure product states.
Consequently, $\rank\r>\max(\rank\r_A,\rank\r_B)$ for any PPTES
$\r$, and any bipartite PPT state of rank $\le3$ is separable.

(ii) If $\r$ is NPT, then it is 1-distillable.

 \epp
We shall apply Proposition \ref{prop:PPTMxNrankN} to the problems of computing the length of separable states, to find the closure of the set of $3\times3$ PPTES of rank four, and to
characterize $2\times N$ separable states studied in Sec.
\ref{sec:Lengths2x3}, \ref{sec:equivalence} and \ref{sec:existence}. So it is an important fact which we use
throughout this paper.

Another useful concept (based on \cite[Definition 11]{cd11JPA})
in this paper is that of reducible and
irreducible states which we are going to introduce now.
 \bd
A linear operator $\r:\cH\to\cH$ is an {\em A-direct sum} of
linear operators $\r_1:\cH\to\cH$ and $\r_2:\cH\to\cH$, written
as  $\r=\r_1\oplus_A\r_2$, if
 $\cR(\r_A)=\cR((\r_1)_A)\oplus\cR((\r_2)_A)$.
 A bipartite state $\r$
 is {\em A-reducible} if it is an A-direct sum of two states;
 otherwise $\r$ is {\em A-irreducible}.
 One defines similarly the {\em B-direct sum} $\r=\r_1\oplus_B\r_2$, the {\em B-reducible} and
 the {\em B-irreducible} states.
 A state $\r$ is {\em reducible} if it is either A or
 B-reducible.
 A state $\r$ is {\em irreducible} if it is both A and B-irreducible.
 \ed

The next result is from \cite[Lemma 15]{cd12}.

\bl
 \label{le:rhoreducible=rhoPTreducible}
Let $\r_1$ and $\r_2$ be linear operators on $\cH$.

(i) If $\r=\r_1\oplus_B\r_2$, then $\r^\G=\r_1^\G\oplus_B\r_2^\G$.

(ii) If $\r_1$ and $\r_2$ are Hermitian and $\r=\r_1\oplus_A\r_2$,
then $\r^\G=\r_1^\G\oplus_A\r_2^\G$.

(iii) If a PPT state $\r$ is reducible, then so is $\r^\G$.
 \el

Let us recall a related result \cite[Corollary 16]{cd11JPA}.
 \bl \label{le:reducible=SUMirreducible,SEP,PPT}
Let $\r=\sum_i\r_i$ be an A or B-direct sum of the states $\r_i$.
Then $\r$ is separable [PPT] if and only if each $\r_i$ is separable
[PPT]. Consequently, $\r$ is a PPTES if and only if each $\r_i$ is
PPT and at least one of them is entangled.
 \el

\section{\label{sec:Lengths2x3}
Lengths of separable states in $2\otimes3$}

We shall need the following result from \cite[Corollary 1, Lemma
2]{kck00}. Their proof is based on their Lemma 1 and is valid for
arbitrary $M,N$. If a Hermitian operator $\r$ is not invertible,
then $\r^{-1}$ will denote its pseudo inverse. (If $\r=\sum_i
p_i\proj{\ps_i}$, $p_i>0$, is the spectral decomposition, then
$\r^{-1}=\sum_i p_i^{-1}\proj{\ps_i}$.)

 \bl
 \label{le:subtractPRODvector}
Let $\r$ be a (non-normalized) bipartite PPT state of birank $(r,s)$
and let $\s=\r-\l\proj{e,f}$ where $\ket{e,f}$ is a product vector
and $\l$ is a real number. Set
$\l_0=(\bra{e,f}\r^{-1}\ket{e,f})^{-1}$ and
$\l_1=(\bra{e^*,f}(\r^\G)^{-1}\ket{e^*,f})^{-1}$. Then $\s$ is a PPT
state if and only if $\ket{e,f}\in\cR(\r)$,
$\ket{e^*,f}\in\cR(\r^\G)$ and $\l\le\min(\l_0,\l_1)$. Moreover, if
$\s$ is a PPT state then its birank is $(r,s)$, $(r-1,s)$, $(r,s-1)$
or $(r-1,s-1)$ according to whether $\l<\min(\l_0,\l_1)$,
$\l=\l_0<\l_1$, $\l=\l_1<\l_0$ or $\l=\l_0=\l_1$.
 \el

Alternatively, this lemma follows from the following simple fact: If
$\r\ge0$ acts on $\cH$, $\l\in\bR$, and $\ket{\phi}\in\cR(\r)$ is a
nonzero vector, then $\r-\l\proj{\phi}\ge0$ if and only if
$\l\bra{\phi}\r^{-1}\ket{\phi}\le1$. Indeed, $\r-\l\proj{\phi}\ge0$
is equivalent to ${\rm id}-\l\r^{-1/2}\proj{\phi}\r^{-1/2}\ge0$. It
remains to observe that $\r^{-1/2}\proj{\phi}\r^{-1/2}$ is a
Hermitian operator of rank one with the nonzero eigenvalue
$\bra{\phi}\r^{-1}\ket{\phi}$.

Next we strengthen part (i) of \cite[Lemma 11]{kck00}.

 \bl \label{le:connected}
$(N\ge M=2)$ Let $V[W]$ be a subspace of the $2\otimes N$ Hilbert
space $\cH$ of dimension $k[l]$ with $k+l>3N$. Then for each unit
vector $\ket{a}\in\cH_A$ there exist infinitely many pairwise
non-parallel unit vectors $\ket{y}\in\cH_B$ such that $\ket{a,y}\in
V$ and $\ket{a^*,y}\in W$. Moreover, the set $S$ of all such pairs
$(\ket{a},\ket{y})$ is connected.
 \el
 \bpf
For the first assertion we essentially follow the proof of
\cite[Lemma 11]{kck00}. Let $f_i$ $(i=1,\ldots,2N-k)$ and $g_j$
$(j=1,\ldots,2N-l)$ be linear functions $\cH\to\bC$ such that
$V=\cap_i \ker f_i$ and $W=\cap_j \ker g_j$. Let $S_A$ $[S_B]$
denote the unit sphere of $\cH_A$ $[\cH_B]$. Let us fix $\ket{a}\in
S_A$. We have $\ket{a,y}\in V$ if and only if $f_i(\ket{a,y})=0$ for
all $i$, and  $\ket{a^*,y}\in W$ if and only if $g_j(\ket{a^*,y})=0$
for all $j$. Since $k+l>3N$, we have $(2N-k)+(2N-l)<N$ and so the
space of solutions of the system of these $4N-k-l$ homogeneous
linear equations for the unknown vector $\ket{y}$ has (complex)
dimension $d_a\ge k+l-3N\ge1$. Hence, the set $S_a$ of all
$\ket{y}\in S_B$ such that $\ket{a,y}\in V$ and $\ket{a^*,y}\in W$
is the unit sphere in some complex subspace of $\cH$ of dimension
$d_a$. In particular, $S_a$ is connected.

Note that $S$ is a closed subset of the product $S_A\times S_B$ and
so it is compact. Let $p_1:S\to S_A$ be the restriction of the first
projection map $S_A\times S_B\to S_A$. We have just shown that $p_1$
is onto and that all of its fibres are connected. This implies that
$S$ itself is connected.
 \epf

We remark that in fact $S$ is a real algebraic subset of $S_A\times
S_B$ and that $\dim S\ge2(k+l-3N)-1$.

From the lemma we deduce an important corollary.

 \bcr \label{cor:birank(r,r)}
Let $\r$ be a $2\times N$ separable state of birank $(r,s)$ with
$r\le s$.

(i) If $r=s$ and $2r>3N$, then there is a product vector $\ket{e,f}$
such that $\s:=\r-\proj{e,f}$ is a PPT state of birank $(r-1,r-1)$.

(ii) If $r<s$ then there is a product vector $\ket{e,f}$ such that
$\s:=\r-\proj{e,f}$ is a PPT state of birank $(r,s-1)$.
 \ecr
 \bpf
We have $\r=\sum^k_{i=1}\proj{a_i,b_i}$ where $k=L(\r)$. The
real-valued function $g$ defined on the set of product vectors by
$g(\ket{e,f})=\bra{e,f}\r^{-1}\ket{e,f}-
\bra{e^*,f}(\r^\G)^{-1}\ket{e^*,f}$ is continuous. Note that $\sum_i
g(\ket{a_i,b_i})=\tr(\r\r^{-1})-\tr(\r^\G(\r^\G)^{-1}) =r-s$.

In case (i) we have $\sum_i g(\ket{a_i,b_i})=0$, and so
$g(\ket{a_i,b_i})\ge0\ge g(\ket{a_j,b_j}$ for some $i$ and $j$. By
Lemma \ref{le:connected}, the set $S$ of normalized product vectors
$\ket{e,f}\in\cR(\r)$ such that $\ket{e^*,f}\in\cR(\r^\G)$ is
connected. Consequently, we have $g(\ket{e,f})=0$ for some product
vector $\ket{e,f}$. The assertion now follows from Lemma
\ref{le:subtractPRODvector} by using this vector and setting
$\l=(\bra{e,f}\r^{-1}\ket{e,f})^{-1}$.

In case (ii) we have $\sum_i g(\ket{a_i,b_i})<0$ and so there exists
an index $i$ such that $g(\ket{a_i,b_i})<0$, i.e.,
$(\bra{a_i,b_i}\r^{-1}\ket{a_i,b_i})^{-1}>
(\bra{a_i^*,b_i}(\r^\G)^{-1}\ket{a_i^*,b_i})^{-1}$. Hence the
assertion  follows from Lemma \ref{le:subtractPRODvector}.
 \epf

 \bpp
 \label{prop:SEP2x3}
If $\r$ is a $2\times3$ separable state of birank $(r,s)$, then
$L(\r)=\max(r,s)$.
 \epp
 \bpf
Without any loss of generality, we may assume that $r\le s$. We
recall that $L(\r)\ge s$ always holds, and that any PPT state in
$2\otimes3$ is separable. By Theorem \ref{thm:PPTMxNrank<M,N}, we
have $r\ge3$.

If $r=3$ then Proposition \ref{prop:PPTMxNrankN} shows that also
$s=3$ and that $L(\r)=3$.

Let $r=4$. If $s=4$ then $L(\r)=4$ by Theorem
\ref{thm:SEP2xN,(N+1,N+1)}. If $s=5$ or 6 we can apply Corollary
\ref{cor:birank(r,r)} (ii) once or twice, respectively, to reduce
these cases to $s=4$.

Let $r=5$. If also $s=5$ then we can apply Corollary
\ref{cor:birank(r,r)} (i) to obtain that $\r=\s+\proj{e,f}$, where
$\s$ is a separable state of birank $(4,4)$. Hence, $L(\s)=4$ and so
$L(\r)=5$. If $s=6$ we can apply Corollary \ref{cor:birank(r,r)}
(ii) to reduce it to the case $s=5$.
 \epf

In Table \ref{tab:SEP2x2}, we recall the well known facts
concerning the lengths of separable $2\times2$ states \cite{stv98,wootters98} (see also \cite[sect III]{hk04}).
Our results concerning the lengths of separable $2\times3$
states are summarized in Table \ref{tab:SEP2x3}. In particular,
note that we have proved that $L(\r)\le6$ for all separable
states on $2\ox3$.
Thus \cite[Conjecture 10]{cd12JPA} is valid in this case.
By inspecting these two tables, it appears that there exist
separable states $\r$ of birank $(r,s)$ when
$\rank\r>\max(\rank\r_A,\rank\r_B)$. In Proposition
\ref{pp:2xN,PPTbirank} below, we shall prove that this is indeed
the case for $2\times N$ separable states. However, it is false
for separable states in general, see Proposition
\ref{pp:PPTES,(N+1+p,N+1+k)}.

 \begin{table}
    \caption{\label{tab:SEP2x2}
Lengths of separable $2\times 2$ states $\r$ of birank $(r,s)$,
$2\le r \le s\le4$. (All such pairs that actually occur are listed.)
Here $\ket{e}=\ket{0}+\ket{1}$.
 } \centering
 \begin{tabular}{|c|c
 |c|c|}
   \hline
   $(r,s)$    & $L(\r)$ & Example &  Reducibility \\\hline
   $(2,2)$    & 2 (see \cite{wootters98}) & $\proj{00}+\proj{11}$ & A,B-reducible \\\hline
   $(3,3)$    & 3 (see \cite{wootters98}) & $\proj{00}+\proj{11}+\proj{e,e}$ & irreducible \\\hline
   $(3,4)$    & 4 (see \cite{wootters98})
   & Example \ref{ex:twoqubit,(34)} & irreducible \\\hline
   $(4,4)$    & 4 (see \cite{wootters98})
   & $I\ox I$ & A,B-reducible \\\hline
 \end{tabular}
 \end{table}

 \begin{table}
    \caption{\label{tab:SEP2x3}
Lengths of separable $2\times 3$ states $\r$ of birank $(r,s)$ with
$3\le r\le s\le6$. (All such pairs that actually occur are listed.)
In the example of birank $(4,6)$, we have $\ket{f}=\ket{0}-\ket{1}$,
$\ket{g}=\ket{0}+\ket{1}+\ket{2}$, $\ket{a_0,b_0}=F[i]$,
$\ket{a_1,b_1}=F[-i]$ where $F[x]:=( (1+x)/(x-1), 1 )^T \ox (-1,
(x-1)/(x+1) , x-1)^T$. Another example of birank $(4,6)$ is
constructed in Example \ref{ex:2x3(4,6)}.
 } \centering
 \begin{tabular}{|c|c|c|c|}
   \hline
   $(r,s)$ & $L(\r)$ & Example &  Reducibility \\\hline
   $(3,3)$ & 3 (see Theorem \ref{prop:PPTMxNrankN}) & $\proj{00}+\proj{11}+\proj{12}$ & A,B-reducible \\\hline
   $(4,4)$ & 4 (see Proposition \ref{prop:SEP2x3}) & $\proj{00}+\proj{01}+\proj{11}+\proj{12}$ &
 A,B-reducible \\\hline
   $(4,5)$ & 5 (see Proposition \ref{prop:SEP2x3}) & Example \ref{ex:TabPrimer} & B-reducible \\\hline
   $(4,6)$ & 6 (see Proposition \ref{prop:SEP2x3}) & $\proj{00}+\proj{11}+\proj{e,2}+\proj{f,g}+\proj{a_0,b_0}+\proj{a_1,b_1}$ & irreducible \\
   \hline
   $(5,5)$ & 5 (see Proposition \ref{prop:SEP2x3})  & $\proj{00}+\proj{01}+\proj{02}+\proj{11}+\proj{12}$ & A,B-reducible \\\hline
   $(5,6)$ & 6 (see Proposition \ref{prop:SEP2x3})  & Example \ref{ex:TabPrimer} & B-reducible \\
   \hline
   $(6,6)$ & 6 (see Proposition \ref{prop:SEP2x3}) & $I\ox I$ & A,B-reducible \\\hline
 \end{tabular}
 \end{table}

\section{\label{sec:equivalence}
Closure of $3\times3$ PPTES of rank four}

The equivalence classes of states are just the orbits under the
action of the group $G=\GL_3(\bC)\times\GL_3(\bC)$. The set, $\cE'$,
of non-normalized $3\times3$ PPTES of rank four is $G$-invariant and
the quotient space $\cE'/G$ parametrizes the set of equivalence
classes of $3\times3$ PPTES of rank four. We equip $\cE'/G$ with the
quotient topology and let $\pi:\cE'\to\cE'/G$ be the projection map.
In this section we shall determine the closure, $\overline{\cE'}$,
of the set $\cE'$ in the ordinary (Euclidean) topology. Note that
the closure, $\overline{\cE}$, of $\cE$ is the intersection
$\overline{\cE'}\cap H$, where $H$ is the space of normalized
Hermitian matrices.

A quantum state $\r$ belongs to the closure, $\overline{\cE'}$,
of the set $\cE'$ if and only if there exist an infinite series of states $\r_1,\r_2,\ldots\in\cE'$ such that
$\lim_{i\ra\infty}\|\r_i-\r\|=0$. So this closure is a set of states attached to the set of two-qutrit PPTES of rank four. The former can be investigated by using the properties of the latter.
We observe that if $\s\in\overline{\cE'}\setminus\cE'$, then $\s$
must be separable and both $\s$ and $\s^\G$ must have rank at
most four. This observation can be used to show that there exist
separable states of rank four which are not in $\overline{\cE'}$. We give an example by modifying \cite[Example 40]{cd12}.

 \bex {\rm \label{ex:TabPrimer}
The separable $2\times3$ state $\s=\proj{00}+\proj{02}+
2\proj{11}+(\ket{01}+\ket{10})(\bra{01}+\bra{10})$ has birank
$(4,5)$. We have $L(\s)=L(\s^\G)\ge\rank\s^\G=5$. Since
$\s-\proj{02}$ is a two-qubit separable state, its length is at most
four \cite{wootters98,stv98}. Hence, $L(\s)$ must be five. As
$\s^\G$ has rank five, $\s\notin\overline{\cE'}$.

Similarly, the separable $2\times3$ state $\s+\proj{12}$ has
birank $(5,6)$ and length six. \hfill $\square$ }
 \eex

On the other hand we have the following result.

 \bl \label{le:sep4}
$(M=N=3)$ We have $\cS'_4\subseteq\overline{\cE}$.
 \el
 \bpf
For convenience, we shall work with non-normalized states. It
suffices to prove that if $\s=\sum_{i=0}^3 \proj{a_i,b_i}$, where
the four product vectors $\ket{a_i,b_i}$ are in general position,
then $\s\in\overline{\cE'}$. Since $\cE'$ and $\overline{\cE'}$ are
$G$-invariant, we may assume that
 \bea
\s=\sum_{i=0}^2 p_i \proj{ii} +\proj{e_A,e_B},
 \eea
where $\ket{e}_A=\sum_i\ket{i}_A$, $\ket{e}_B=\sum_i\ket{i}_B$ and
the $p_i$ are positive scalars.

We consider the states $\r=\r(a,b,d)=C^\dag C$, where
$C=[C_0~C_1~C_2]$ and the blocks $C_i$ are $4\times3$ matrices in
Eq. (\ref{ea:Blokovi}) with $c=0$. Clearly, $\r$ belongs to the
closure of $\cE'$. It is easy to verify that $\r=\sum_{i=0}^3 p_i
\proj{v_i}$, where
 \bea
&& p_0=\frac{1}{1+b^2},\quad p_1=\frac{1}{1+d^2},\quad
p_2=\frac{1}{d^2(1+d^2)},\quad p_3=\frac{1}{b^2(1+b^2)}; \\
&& \ket{v_0}=\ket{0}\ox(ab\ket{1}+(1+b^2)\ket{2}), \\
&& \ket{v_1}=(d\ket{1}+(1+d^2)\ket{2})\ox\ket{0}, \\
&& \ket{v_2}=\ket{1}\ox(d\ket{0}-(1+d^2)\ket{2}), \\
&& \ket{v_3}=(ab\ket{0}-(1+b^2)\ket{2})\ox\ket{1}.
 \eea
Let $V=b(1+b^2)^{-3/2}V_A\ox V_B$ where
 \bea
V_A=\left[ \begin{array}{ccc}
(1+b^2)/ab & 0 & 0 \\
0 & 0 & -1 \\
0 & (1+d^2)/d & -1 \end{array} \right], \quad V_B=\left[
\begin{array}{ccc}
0 & 1+b^2 & 0 \\
-ab(1+d^2)/d & 1+b^2 & -ab \\
0 & 1+b^2 & -ab \end{array} \right].
 \eea
A computation shows that $V\r V^\dag=\s$ provided we choose the
positive parameters $a,b,d$ such that
 \bea
b^2=p_0, \quad \frac{a^2b^4}{d^2}\cdot
\left(\frac{1+d^2}{1+b^2}\right)^3=p_1, \quad d^2=\frac{p_1}{p_2}.
 \eea
 \epf

We can now show that $\overline{\cE'}$ contains many separable
states.

 \bl
 \label{le:SEPrank3}
Separable states of rank at most three have length at most four.
 \el
 \bpf
Let $\r$ be a separable $k\times l$ state of rank $r\le3$. We may
assume that $k\le l$. By \cite[Theorem 1]{hst99}, we have $l\le r$.
The assertion is trivial if $l=1$, it follows from
\cite{wootters98,stv98} if $l=2$, and from \cite[Proposition
9]{cd12} if $l=3$.
 \epf

 \bl
 \label{le:3x3rank4le5}
The maximum length of $3\times3$ separable states of rank four is
five.
 \el
 \bpf
Separable $3\times3$ states of rank four and length five exist, see
e.g. \cite[Example 40]{cd12}. Let $\r$ be any $3\times3$ separable
state of rank four and length $r>4$. Thus we have
$\r=\sum^{r-1}_{i=0}\proj{a_i,b_i}$. We may assume that the
$\ket{a_i,b_i}$ with $i<4$ are linearly independent. By \cite[Lemma
29]{cd12}, these four product vectors are not in general position.
Consequently, we may assume that  $\ket{b_0}=\ket{0}$,
$\ket{b_1}=\ket{1}$, $\ket{b_3}=\ket{2}$ and $\braket{b_2}{2}=0$.
Moreover, we may assume that $\ket{a_i}=\ket{a_3}$ for $3\le i\le
s<r$, while for $i>s$ the vectors $\ket{a_i}$ are not parallel to
$\ket{a_3}$. It is not hard to show that we can rewrite
$\sum^s_{i=3}\proj{b_i}$ as $\proj{b_3'}+\s$, where $\s$ is a state
on $\cH_B$ such that $\s\ket{2}=0$. Clearly, we have
$\braket{b'_3}{2}\ne0$ and so $\cR(\r)$ is spanned by
$\ket{a_i,b_i}$, $i=0,1,2$ and $\ket{a_3,b_3'}$. Since
$\ket{a_i,b_i}\in\cR(\r)$, it follows that for $i>s$ we must have
$\braket{b_i}{2}=0$. Consequently, we have a B-direct decomposition
$\r=\r'\op_B\proj{a_3,b_3'}$. Since $\r'$ is separable of rank
three, its length is at most four by Lemma \ref{le:SEPrank3}. Hence
$\r$ has length five.
 \epf

From the lemma we obtain
 \bcr
 \label{cr:3x3rank4=CAR5}
A $3\times3$ separable state $\r$ of rank four has length five if
and only if it is A or B-direct sum of a pure product state and a
separable state $\s$ of rank three and length four.
 \ecr
 \bpf
\textit{Necessity}. See the proof of Lemma \ref{le:3x3rank4le5}.

\textit{Sufficiency}. Suppose that $\r=\s\op_B\proj{a,b}$, with $\s$
a separable state of rank three and length four. As length does not
increase under local operations, we have $L(\r)\ge L(\s)=4$. Assume
that $L(\r)=4$ and so
$\r=\sum^3_{i=0}\proj{a_i,b_i}=\s\op_B\proj{a,b}$. Suppose
$\braket{b}{b_i}\ne0$ for $i=0,\cdots,s$. Then for these subscripts
$\ket{a_i}$ are pairwise parallel, and we may assume
$\braket{b}{b_i}\ne0$ for only $i=0$. Thus $\ket{b_0}$ is
proportional to $\ket{b}$. The equality
$\sum^3_{i=0}\proj{a_i,b_i}=\s\op_B\proj{a,b}$ indicates
$\rank\s=3$, which gives us a contradiction. This completes the
proof.
 \epf

 \bl \label{le:SEP(4,4)}
A $3\times3$ separable state has birank $(4,4)$ if and only if it
has length four.
 \el
 \bpf
\textit{Necessity}. Suppose $\r$ is a $3\times3$ separable state of
birank $(4,4)$. By Lemma \ref{le:3x3rank4le5}, $L(\r)\le5$. Assume
that $L(\r)=5$. By using Corollary \ref{cr:3x3rank4=CAR5} we obtain
that, say, $\r=\s\op_A\proj{a,b}$ where $\s$ is a $2\times2$ or
$2\times3$ separable state of rank three and length four. It follows
from Proposition \ref{prop:PPTMxNrankN} (a) that $\s$ must be
$2\time2$ state. From Table \ref{tab:SEP2x2}, we see that
$\rank\s^\G=4$. By Lemma \ref{le:rhoreducible=rhoPTreducible} (ii),
we have $\r^\G=\s^\G\op_A\proj{a^*,b}$. Therefore $\rank\r^\G=5$,
which gives a contradiction. So $\r$ must have length four.

\textit{Sufficiency}. Suppose $\r$ is a $3\times3$ separable state
of length four. Suppose its birank is $(r,s)$, then $4\ge r,s\ge3$.
If either of $r,s$ is equal to three, then $L(\r)=3$ by using
Proposition \ref{prop:PPTMxNrankN}. It gives us a contradiction, so
$r=s=4$.
 \epf

We can now prove the main result of this section.
 \bt \label{thm:Closure}
$(M=N=3)$ We have $\overline{\cE}=\cE\cup\cS'_4$.
 \et
 \bpf
Let $\r\in\overline{\cE}$ be separable. Then $\r$ is a $k\times l$
state of birank $(r,s)$ with $\max(r,s)\le4$. In view of Lemma
\ref{le:sep4}, it suffices to prove that $L(\r)\le4$. Recall that
$L(\r)=L(\r^\G)$. If $r<4$ or $s<4$ then $L(\r)\le4$ by Lemma
\ref{le:SEPrank3}. Assume now that $r=s=4$. If $k=l=3$ then
$L(\r)\le4$ by Lemma \ref{le:SEP(4,4)}. If $(k,l)$ is equal to
$(2,3)$ or $(3,2)$, then $L(\r)=4$ by Proposition \ref{prop:SEP2x3}.
Otherwise, $k=l=2$ and $L(\r)\le4$ by \cite{dtt00}. Hence, the proof
is completed.
 \epf

Recall that any $\r\in\cE'$ is equivalent to $\r^\G$ \cite[Theorem
23]{cd11JMP}. The following example shows that this property does
not extend to $\overline{\cE'}$.

 \bex
 \label{ex:twoqubit,(34)} {\rm
The separable $2\times2$ state
$\s=2\proj{00}+\proj{11}+(\ket{01}+\ket{10})(\bra{01}+\bra{10})$ has
birank $(3,4)$, and so $\s$ is not equivalent to $\s^\G$. On the
other hand, since $L(\s)=4$, we have $\s\in\overline{\cE'}$ by Lemma
\ref{le:sep4}. Explicitly, we have
 \bea
 \s &=& \proj{00} + \frac13 \left( \proj{\ps_0}
 + \proj{\ps_1} + \proj{\ps_2} \right), \\
\ket{\ps_k}&=&(\ket{0}+\z^k\ket{1})\ox(\ket{0}+\z^k\ket{1}), \quad
k=0,1,2;
 \eea
where $\z:=(-1+i\sqrt{3})/2$ is a primitive cube root of unity.

We can now show that the quotient space $\overline{\cE'}/G$ is not
Hausdorff. Indeed, let $(\r_i)$ be a sequence in $\cE'$ converging
to $\s$. Then the sequence $(\r_i^\G)$ converges to $\s^\G$.
Consequently, the sequence $(G\cdot\r_i)$ converges to $G\cdot\s$
and the sequence $(G\cdot\r_i^\G)$ converges to $G\cdot\s^\G$ in
the space $\overline{\cE'}/G$. But these two sequences coincide
because $\r_i^\G$ is equivalent to $\r_i$ for each $i$. On the
other hand, the points $G\cdot\s$ and $G\cdot\s^\G$ are distinct
because the states $\s^\G$ and $\s$ are not equivalent (they have
different ranks). Hence, the sequence $(G\cdot\r_i)$ converges to
two different points and we conclude that the space
$\overline{\cE'}/G$ is not Hausdorff.
\hfill $\square$ }
 \eex

Finally we propose an application of two-qutrit PPTES of rank four.
Consider a separable state $\r$ of birank $(r,s)$, and the set $S$
of product vectors $\ket{e,f}\in\cR(\r)$ and
$\ket{e^*,f}\in\cR(\r^\G)$, such that $\s=\r-\proj{e,f}$ is a PPT
state of birank equal to $(r-1,s)$, $(r,s-1)$, or $(r-1,s-1)$. We
are going to construct a family of $\r$ such that any $\s$ is PPTES.

 \bex
 \label{ex:3x3pptes}
 $(M=N=3)$ {\rm
Let $\r$ be a $3\times3$ PPTES of rank four. Then $\ker\r$ contains
exactly six product vectors (up to a scalar factor) $\ket{\psi_i}$,
$i=1,\ldots,6$, and moreover any five of these vectors are linearly
independent, see Ref. \cite{cd11JMP}. Consequently, the six rank-one
operators $\proj{\ps_i}$ are linearly independent. Since $\r^\G$ is
also a $3\times3$ PPTES of rank four, the partial conjugates of the
$\ket{\psi_i}$ have similar properties.

We consider the separable state
 \bea
 \label{ea:3x3UPBsep}
\s = \sum^6_{i=1} \proj{\ps_i}
 \eea
of birank $(5,5)$. Let $\ket{e,f}$ be a product vector such that
$\s':=\s-\proj{e,f}$ is a PPT state of birank $(r,s)$ with $r\le5$,
$s\le5$ and $r+s<10$. (By Lemma \ref{le:subtractPRODvector}, we know
that such product vector exists.) By the same lemma,  $\proj{e,f}$
must be a scalar multiple of some $\proj{\psi_i}$, say
$\proj{e,f}=c\proj{\psi_1}$. Clearly, we must have $c>1$.

We claim that $\s'$ must be entangled. Indeed, if $\s'$ is
separable, then it can be written as $\s'=\sum_i c_i \proj{\ps_i}$
with $c_i\ge0$. Since the $\proj{\ps_i}$ are linearly independent,
it follows that $c_1=1-c$. Hence, $c=1-c_1\le1$ which gives a
contradiction.
 \hfill $\square$ }
 \eex

We do not know that whether there is a similar example in
$2\ox4$. The following lemma is evident. It implies that the
length of the state \eqref{ea:3x3UPBsep} is six.
 \bl
Let $\r$ be a separable state with $\rank\r=L(\r)=r$. Then there
is a product vector $\ket{a,b}$, such that $\s:=\r-\proj{a,b}$
is a separable state with $\rank\s=L(\s)=r-1$.
 \el

\section{\label{sec:existence}
qubit-qudit PPT states with prescribed birank}

So far we have mainly focused on $2\times3$ and $3\times3$ PPT
states. In this section we investigate some typical types of
$2\times N$ PPT states $\r$ for arbitrary $N$. In Theorem
\ref{thm:SEP2xN,(N+1,N+1)} we characterize both separable and PPT
entangled states $\r$ of birank $(N+1,N+1)$. This case is different
from those discussed in Corollary \ref{cor:birank(r,r)}. In
Proposition \ref{pp:Prva} we study the properties of the set of
product vectors contained in $V^\perp$, where $V$ is a CES of
dimension $k<N$ in $2\ox N$. It turns out that this set (with zero
vectors included) is a vector bundle of rank $N-k$ over the
projective line $\cP^1$. In the special case $k=N-1$, the
projectivization of this set is a rational normal curve. In
Propositions \ref{pp:2xN,PPTbirank} and \ref{pp:PPTES,(N+1+p,N+1+k)}
we construct separable states and PPTES of any birank $(r,s)$ with
$r,s>N$. The constructions are based on Proposition \ref{pp:Prva}
and the recently constructed PPTES in \cite{tah12}. Finally we
obtain a result on NPT states. In Example
\ref{ex:negativeEIGENVALUE}, for each $m=1,\ldots,N-1$, we construct
$2\times N$ NPT state whose partial transpose has exactly $m$
negative eigenvalues.

A PPT state $\r$ is an \textit{edge state} if there is no product
vector $\ket{a,b}\in\cR(\r)$ such that
$\ket{a^*,b}\in\cR(\r^\G)$. Any edge state is necessarily entangled. Any bipartite PPTES is the sum of a separable state
and an edge state \cite{lkc00}. So, in the bipartite case, edge
states play the role of ``extreme points'' in the set of PPTES.
It is useful to describe the structure of states in the following family.

 \bt
 \label{thm:SEP2xN,(N+1,N+1)}
Let $\r$ be a $2\times N$ PPT state of birank $(N+1,N+1)$.

(i) If $\r$ is separable then $L(\r)=N+1$.

(ii) If $\r$ is entangled then
$\r=\s\op_B\proj{a_1,b_1}\op_B\cdots\op_B\proj{a_r,b_r}$, where
$\s$ is an edge state of birank $(N+1-r,N+1-r)$.
 \et
 \bpf
(i) First note that $L(\r)\ge\rank\r=N+1$. Table \ref{tab:SEP2x2}
shows that the assertion is true for $N=2$. We proceed by
induction on $N$. Now let $N>2$. Since $\r$ is separable, by
Lemma \ref{le:subtractPRODvector} we have $\r=\s+\proj{e,f}$
where $\s$ is a PPT state of birank $(N,N+1)$, $(N+1,N)$ or
$(N,N)$, and $\ket{e,f}$ is a product vector. If $\rank\s_A=1$, the assertion
clearly holds, and so we may assume that $\rank\s_A=2$. Since
$\r_B=\s_B+\|e\|^2\proj{f}$, we have $\rank\s_B=N$ or $N-1$. If
$\rank\s_B=N$, the assertion follows from Proposition
\ref{prop:PPTMxNrankN}. Otherwise, $\rank\s_B=N-1$ and Lemma
\ref{le:reducible=SUMirreducible,SEP,PPT} shows that $\s$ is
separable of birank $(N,N)$. By the induction hypothesis,
$L(\s)=N$ and consequently $L(\r)=N+1$.

(ii) If $\r$ is an edge state, then the assertion holds with
$r=0$. Otherwise, by Lemma \ref{le:subtractPRODvector}, we have
$\r=\s+\proj{e,f}$, where $\s$ is a PPT state of birank
$(N,N+1)$, $(N+1,N)$ or $(N,N)$, and $\ket{e,f}$ is a product
vector. As $\r$ is entangled, we must have $\rank\s_A=2$. We also
have $\rank\s_B=N$ or $N-1$. Proposition \ref{prop:PPTMxNrankN}
implies that $\rank\s_B=N-1$, and so $\r=\s\op_B\proj{e,f}$. By
Lemma \ref{le:rhoreducible=rhoPTreducible} (i), $\s$ has birank
$(N,N)$. We can continue to apply this procedure of splitting off
a pure product state as long as the entangled summand is not an
edge state. Eventually, this summand must become an edge state.
This completes the proof.
 \epf

We point out that part (i) generalizes the $2\ox3$ case in Table
\ref{tab:SEP2x3}, and that part (ii) was also discussed in
\cite[Sec. IV B]{kck00}. We further point out that $M\times N$ PPT
states $\r$, with $N\ge M\ge3)$, of rank $N+1$ have been
investigated in \cite[Theorems 44, 45]{cd12}. In particular, the
first of these theorems implies that
$\r=\r_1\oplus_B\cdots\oplus_B\r_k\oplus_B\s$, where $\r_i$ are pure
product states and $\s$ is a B-irreducible state. Note that this
decomposition is similar to one in Theorem
\ref{thm:SEP2xN,(N+1,N+1)} (ii). In physics, such a decomposition
means that the entanglement of $\r$ is "absolutely" robust to the
noise of separable states
$\a=\proj{a_1,b_1}\op_B\cdots\op_B\proj{a_r,b_r}$ in the following
sense: the normalized state $\r=(1-p)\s+p\a$ is always entangled no
matter how big the weight $p<1$ is. This phenomenon usually does not
occur for other $2\times N$ entangled states, which would become
separable by adding a separable state.

It was proved recently \cite[Theorem 5]{fg10} that in $2\ox N$
the PPT states of birank $(2N,k)$ exist if and only if
$N<k\le2N$. We shall obtain another existence result which, in
particular, shows that there exist $2\times N$ separable states  of birank $(N+j,N+k)$ for any $j,k=1,\ldots,N$.
For that we need two lemmas proved in
\cite[Lemmas 1,2]{atl11}. In the next proposition we give a novel
proof of the strengthened version of the combination of these
two lemmas. For the definition and basic properties of the
rational normal curves used in this lemma, see
\cite[p. 10-14]{h92}.

\bpp \label{pp:Prva}
We consider the bipartite system $2\ox N$ with Hilbert space
$\cH=\cH_A\ox\cH_B$ of dimension $2N$.
Let $V\sue\cH$ be a CES of dimension $k<N$ and let $Y$ be the set of all product vectors in $V^\perp$.

(i) The set $Y$ (with zero vectors included) is an algebraic
vector bundle of rank $N-k$ over the projective line.

(ii) $V^\perp$ is spanned by $Y$.

(iii) The partial conjugates of members of $Y$ span the whole space $\cH$.

(iv)  If $k=N-1$ the projectivization of $Y$ is a rational normal
curve.
\epp
\bpf
Let $\ket{\psi_i}=\ket{0}\ox\ket{a_i}+\ket{1}\ox\ket{b_i}$,
$i=1,\ldots,k$, be a basis of $V$. We introduce the
$2\times N$ matrices
\bea \label{eq:MatriceR}
R_i=\left[ \begin{array}{cccc}
\a_{i0} & \a_{i1} & \cdots & \a_{i,N-1} \\
\b_{i0} & \b_{i1} & \cdots & \b_{i,N-1} \end{array} \right]^*,
\quad i=1,\ldots,k,
\eea
where $\sum_j \a_{ij}\ket{j}=\ket{a_i}$ and
$\sum_j \b_{ij}\ket{j}=\ket{b_i}$.
Since $V$ is a CES, if the scalars $\xi_i$, $i=1,\ldots,k$,
are not all zero then
\bea \label{eq:Rang2}
\rank \sum_{i=1}^k \xi_i R_i = 2.
\eea

The projectivization of $\cH_A$ is a projective line $\cP^1$.
The point of $\cP^1$ corresponding to the nonzero vector
$z\ket{0}+w\ket{1}\in\cH_A$ will be denoted by $[z:w]$.
We claim that for each point $[z:w]\in\cP^1$, the set
of all vectors $\ket{f}\in\cH_B$ such that
$(z\ket{0}+w\ket{1})\ox\ket{f}\in V^\perp$
is a vector subspace of dimension $N-k$.
We shall use the expansion
$\ket{f}=\sum_j f_j \ket{j}\in\bC^N$, $f_j\in\bC$.
To find the coefficients $f_j$ we have to solve the system
of $k$ linear homogeneous equations
$\bra{\psi_i}(z\ket{0}+w\ket{1})\ox\ket{f}=0$, i.e.,
\bea \label{eq:Sistem}
\sum_{j=0}^{N-1} (\a_{ij}^*z+\b_{ij}^*w) f_j=0,\quad
i=1,\ldots,k,
\eea
with matrix $C$ of size $k\times N$.
Suppose that for some $x=(\xi_1,\ldots,\xi_k)\in\bC^k$
we have $xC=0$. We can rewrite this equation as
$(z,w)\cdot\sum_i \xi_i R_i=0$. Eq. \eqref{eq:Rang2} implies
that $x=0$, and so $\rank C=k$. Consequently, the set of
solutions of the system \eqref{eq:Sistem} is a vector space of
dimension $N-k$, and the claim is proved.
Thus the fibres of the projection map $p:Y\to\cP^1$ are vector
spaces of dimension $N-k$, and (i) follows.

The matrix $C$ is in fact a matrix pencil $C=Az+Bw$, where
$A=[\a_{ij}^*]$ and $B=[\b_{ij}^*]$ are $k\times N$ complex
matrices. We shall use the Kronecker's theory of matrix pencils
as presented in the well known book of Gantmacher \cite{g89}.
He writes a matrix pencil in non-homogeneous form as
$A+\lambda B$, where $\lambda$ is an indeterminate.
We homogenize the notation by setting $\lambda=w/z$ and multiplying the pencil by $z$.
The canonical form for matrix pencils is a direct sum of blocks
of several types: $L_m$, their transposes $L_m^T$, $N^{(u)}$,
and $wJ+zI_s$ where $I_s$ is the identity matrix of order $s$
and $J$ a Jordan block. As we shall see below, it turns out
that we have to deal only with the blocks of type
\bea
L_m=\left[ \begin{array}{cccccc}
z&-w&0&\cdots&0&0\\
0&z&-w&      &0&0\\
\vdots&&&&&\\
0&0&0&      &z&-w \end{array} \right],
\eea
of size $m\times(m+1)$. To simplify notation in some formulae
below, we have replaced $w$ with $-w$ which we can obviously do.
For instance, we have
\bea
\left[\begin{array}{ccc}1&0&0\\0&-1&0\\0&0&1\end{array}\right]
\cdot
\left[\begin{array}{cccc}z&w&0&0\\0&z&w&0\\0&0&z&w\end{array}
\right]\cdot
\left[\begin{array}{cccc}1&0&0&0\\0&-1&0&0\\0&0&1&0\\0&0&0&-1
\end{array}\right]=
\left[\begin{array}{cccc}z&-w&0&0\\0&z&-w&0\\0&0&z&-w
\end{array}\right].
\eea
Contrary to Gantmacher, we allow the index $m$ of the block $L_m$ to be 0 in which case $L_m$ has size $0\times1$.
There exist invertible matrices $P$ and $Q$ (whose entries are
complex constants independent of $z$ and $w$) such that
$C':=PCQ$ has the canonical form given by
\cite[Eq. (34), p. 39]{g89}.
By changing the basis of $\cH_A$, we may assume that $Q=I_N$
is the identity matrix.
Any row of $C'$ has the form $(z,w)\cdot\sum_i \xi_i R_i$,
where $\xi_i\in\bC$ are some constants, not all 0.
Hence, the rank condition \eqref{eq:Rang2} implies that each
row of $C'$ must have at least two nonzero entries.
This is a very strong condition, it implies that $C'$ consists
only of blocks of type $L_m$. Since $L_m$ has size
$m\times(m+1)$, there are exactly $N-k$ blocks, i.e., we have
\bea
C'=L_{m_1}\oplus\cdots\oplus L_{m_{N-k}},
\eea
where $m_1+\cdots+m_{N-k}=k$.
Consequently, the system \eqref{eq:Sistem} breaks up into
$N-k$ simple independent subsystems of linear homogeneous
equations $L_{m_i}f^{(i)}=0$, $i=1,\ldots,N-k$. For instance,
the first subsystem comprises only the unknowns
$f_1,\ldots,f_{m_1+1}$
which are the components of the column vector $f^{(1)}$, etc.
Since $L_{m_i}$ has rank $m_i$, the $i$th subsystem has a
unique solution (up to a scalar factor) when viewed as a
system of equations in its own portion of the unknowns $f_j$.
There is a unique solution whose unknowns are just monomials
in $z$ and $w$ of total degree $m_i$. We refer to this solution
as the {\em basic solution}. For instance, for the first subsystem the basic solution is given by
\bea
f_1=w^{m_1},~ f_2=zw^{m_1-1},~\ldots~,~f_{m_1+1}=z^{m_1}.
\eea
Note that if $m_1=0$ then the first subsystem has only
one unknown, namely $f_1$, but it has no equations. The
basic solution in that case is just $f_1=1$.
For convenience, we shall identify this basic solution
with the vector
$\ket{g^{(1)}}=\sum_{i=0}^{m_1} z^i w^{m_1-i} \ket{i}\in\cH_B$.
The other subsystems can be solved in the same manner.
Their basic solutions are given explicitly by
\bea \label{eq:Basic}
\ket{g^{(i)}}=\sum_{j=0}^{m_i} z^jw^{m_i-j}\ket{m'_{i-1}+j},
\quad i=1,\ldots,N-k,
\eea
where $m'_{i-1}=m_1+\cdots+m_{i-1}+i-1$ (with $m'_0=0$).
The general solution is given by an arbitrary linear combination
of the basic solutions $\ket{g^{(i)}}$, $i=1,\ldots,N-k$.
We shall form a special solution in which the coefficients
of this linear combination are suitably chosen monomials in
$z$ and $w$. Thus we shall multiply $g^{(i)}$ with some
monomial $z^{u_i}w^{v_i}$. After expanding the tensor product
$(z\ket{0}+w\ket{1})\ox z^{u_1}w^{v_1}\sum_{j=0}^{m_1}
z^j w^{m_1-j}\ket{j}$,
we obtain a linear combination of the basis vectors
with $m_1+2$ different monomial coefficients
$z^{u_1+j+1}w^{v_1+m_1-j}$ with $j=-1,0,1,\ldots,m_1$.
We can choose the exponents $u_i,v_i$ so that the monomials
arising from different subsystems are all different and
moreover the total degree $\d:=m_i+u_i+v_i$ is independent
of the index $i$. Then the total number of different monomials
that occur in the expansion of
\bea \label{eq:ProdVekt}
&&(z\ket{0}+w\ket{1})\ox \sum_{i=1}^{N-k}
z^{u_i}w^{v_i}\ket{g^{(i)}}
\eea
is $\sum_{i=1}^{N-k} (m_i+2)=2N-k$. Since these $2N-k$
monomials are linearly independent, we conclude that the
product vectors \eqref{eq:ProdVekt} span a subspace of
dimension $2N-k$. Since all of them belong to $V^\perp$,
the assertion (ii) is proved.


The assertion (iii) follows by using a similar argument as
above after replacing
$z\ket{0}+w\ket{1}$ with $z^*\ket{0}+w^*\ket{1}$ and
observing that the $2N$ ``monomials''
$z^*z^{u_i+j}w^{v_i+m_i-j},w^*z^{u_i+j}w^{v_i+m_i-j}$,
where $i=1,\ldots,N-k$ and for fixed $i$ the index $j$ takes the values $0,1,\ldots,m_i$, are linearly independent.
Indeed, any nontrivial linear dependence relation among these ``monomials'' would give an identity $z^*p(z,w)+w^*q(z,w)=0$,
where $p(z,w)$ and $q(z,w)$ are nonzero homogeneous polynomials
in $z$ and $w$ of degree $\d$. By dehomogenizing, i.e.,
dividing this identity by $z^*z^\d$, we obtain that $(w/z)^*$
is an analytic function of $w/z$, which is a contradiction.

In the case $k=N-1$, we have $C'=L_{N-1}$ and so all product
vectors in $V^\perp$ have the form $(z\ket{0}+w\ket{1})\ox
\sum_{i=0}^{N-1} z^{N-1-i}w^i \ket{i}$.
The assertion (iv) follows.
 \epf

Note that Theorem \ref{thm:PPTMxNrank<M,N} implies that if
$(r,s)$ is a birank of a $2\times N$ PPT state then $r,s\ge N$,
and Proposition \ref{prop:PPTMxNrankN} shows that $r=N$ if and only if $s=N$.
Now we can show that, for any $r,s\in\{N+1,\ldots,2N\}$,
there exist $2\times N$ separable states of birank $(r,s)$.

 \bpp
 \label{pp:2xN,PPTbirank}
There exist $2\times N$ separable states of birank $(N+j,N+k)$
for any $j,k\in\{1,\ldots,N\}$.
 \epp
 \bpf
The identity operator on $\cH$ is a separable state of birank
$(2N,2N)$. Thus, we may assume that $j\le k\le N$ and $j<N$.
Let $V$ be a CES of dimension $N-j$. By Proposition
\ref{pp:Prva} (ii), $V^\perp$ has a basis consisting of product
vectors, say $\ket{e_i,f_i}$, $i=1,\ldots,N+j$. The space $W$
spanned by their partial conjugates has dimension at most $N+j$.
By Proposition \ref{pp:Prva} (iii), there exist
product vectors $\ket{e'_s,f'_s}\in V^\perp$, $s=1,\ldots,m$,
such that the partial conjugates of the $\ket{e_i,f_i}$ and
the $\ket{e'_s,f'_s}$ together span a space $W'\supseteq W$ of dimension $N+k$. Then the sum of all $\proj{e_i,f_i}$ and all
$\proj{e'_s,f'_s}$ is a separable state of birank $(N+j,N+k)$.
 \epf

(According to the authors of \cite{ck12}, this proposition is
contained in Sect. III of their paper.)

Let us give an {\em ad hoc} example for the case $N=3$ with
$(r,s)=(4,6)$.

 \bex
 \label{ex:2x3(4,6)} {\rm
We have constructed an explicit separable $2\times3$ state $\r$ of
birank $(4,6)$ and length six. It can be written as $\r=\sum_{i=1}^4
\proj{\psi_i}$ where
 \bea
\ket{\psi_1} &=& 2\ket{00}, \\
\ket{\psi_2} &=& \ket{1}(\ket{0}+2\ket{1}), \\
\ket{\psi_3} &=& 2\ket{01}+(\ket{0}+\ket{1})\ket{2}, \\
\ket{\psi_4} &=& \ket{02}+\ket{1}(\ket{0}-\ket{1}-\ket{2}).
 \eea
Since the characteristic polynomial of $\r^\G$ is
$t^6-19t^5+133t^4-413t^3+520t^2-148t+4$, we have $\r^\G>0$.
Consequently, $\r$ is separable of birank $(4,6)$. By Lemma
\ref{prop:SEP2x3}, $\r$ has length six. \hfill $\square$ }
 \eex

Let us now show that there exist $2\times N$ PPTES of birank
$(N+1,N+k)$ for $k=1,\ldots,N$. We shall do that by using
a recently constructed family \cite[Eq. (5),Appendix B]{tah12}
of $2\times N$ PPTES of birank $(N+1,N+1)$.
By dropping the normalization and setting the parameter
$b=1/2$, we obtain the $2\times N$ PPTES
 \bea \label{eq:Stanje}
 \r
 &:=&
 \sum^{N-2}_{i=0}(\ket{0,i}+\ket{1,i+1})(\bra{0,i}+\bra{1,i+1}) +\proj{10}
 \notag\\
 &+&
 \frac12\ket{0}(\ket{0}+\sqrt{3}\ket{N-1})
        \bra{0}(\bra{0}+\sqrt{3}\bra{N-1}).
 \eea
Its partial transpose is
 \bea
 \r^\G
 &=&
 \sum^{N-2}_{i=0} (\ket{0,i+1}+\ket{1,i})(\bra{0,i+1}+\bra{1,i}) +\proj{1,N-1}
 \notag\\
 &+&
 \frac12\ket{0}(\sqrt{3}\ket{0}+\ket{N-1})\bra{0}(\sqrt{3}\bra{0}+\bra{N-1}).
 \eea
One can verify that
 \bea
 \ket{\ph(a)}:=(\ket{0}+a\ket{1})
 ((a^{N-1}+\frac{1}{\sqrt{3}})\ket{0}+a^{N-2}\ket{1}+\cdots+a\ket{N-2}+\ket{N-1})\in \cR(\r)
 \eea
for all $a\in\bC$, and that the $\ket{\ph(a)}$ with $a\in\bR$
span $\cR(\r)$.

 \bl
 \label{le:PPTES,(N+1,N+1+p)}
For sufficiently small $\e>0$ and $k\in\{1,\ldots,N-1\}$, the
state $\r_k:=\r+\e\sum^k_{i=1} \proj{\ph(a_i)}$ is a
$2\times N$ PPTES of birank $(N+1,N+1+k)$.
 \el
 \bpf
Since $\e>0$ is small and $\r$ is a $2\times N$ PPTES, so is
$\r_k$. Since $\ket{\ph(a)}\in\cR(\r)$,
it follows that $\rank\r_k=N+1$. One can verify that
$\cR(\r)+\cR(\r^\G)=\cH$. Hence,
there are distinct real numbers $a_i$, $i=1,\ldots,N-1$, such
that the vectors $\ket{\ph(a_i)}$ are linearly independent
modulo $\cR(\r^\G)$. Since the $a_i$ are real,
each product vector $\ket{\ph(a_i)}$ is equal to its partial conjugate. It follows that $\rank\r_k^\G=N+1+k$.
 \epf

More generally, we have the following result.
 \bpp
\label{pp:PPTES,(N+1+p,N+1+k)}
For any $r,s\in\{N+1,\ldots,2N\}$, there exist $2\times N$ PPTES
of birank $(r,s)$.
 \epp
 \bpf
Let $k,p\in\{0,\ldots,N-1\}$ and let $\r_k$ be the state
constructed in Lemma \ref{le:PPTES,(N+1,N+1+p)}.
For the state $\r$ defined by Eq. \eqref{eq:Stanje}, we have
$\r^\G=(I\ox V)\r(I\ox V^\dg)$ where $V$ is the
anti-diagonal matrix. So $\cR(\r^\G)$ is spanned by the
product vectors $\ket{\ps(a)}=(I\ox V)\ket{\ph(a)}$ with
$a\in\bR$.
Since $\cR(\r)+\cR(\r^\G)=\cH$, there are distinct real numbers
$a'_j$, $j=1,\ldots,N-1$, such that the product vectors
$\ket{\ps(a'_j)}$ are linearly independent modulo $\cR(\r)$.
Note that $\proj{\ps(a'_i)}^\G=\proj{\ps(a'_i)}$ for each $i$.
It follows that, for sufficiently small $\e'>0$, the state
$\r_k+\e'\sum^p_{i=1} \proj{\ps(a'_i)}$ is a $2\times N$ PPTES
of birank $(N+1+p,N+1+k)$.
 \epf

One may expect that Propositions \ref{pp:2xN,PPTbirank} and
\ref{pp:PPTES,(N+1+p,N+1+k)} generalize to arbitrary $M\ox N$ space, i.e., that $M\times N$ separable states as well as PPTES
of birank $(r,s)$ exist for all $r,s>\max(M,N)$. However, this
is false. For the former, we observe that there is no separable
$3\times3$ state of birank $(4,6)$. Indeed, let $\r$ be any
$3\times3$ separable state of rank four. By Lemma
\ref{le:3x3rank4le5}, $\rank\r^\G\le5$. Then Proposition
\ref{prop:PPTMxNrankN} (i) implies that $\rank\r^\G<6$. For the
latter, we observe that there is no two-qutrit PPTES of birank $(4,5)$ or $(4,6)$ (see \cite[Theorem 23]{cd11JPA}).

We give a result on NPT states as the concluding remark of this
section. It has been shown that, for any NPT $2\times N$ state, its partial transpose has at most $N-1$ negative eigenvalues
\cite[Theorem 1]{rp12}. This upper bound is sharp. More
precisely, for each $m\in\{1,\ldots,N-1\}$, we shall construct
$2\times N$ NPT states whose partial transpose has exactly $m$
negative eigenvalues.

 \bex
 \label{ex:negativeEIGENVALUE}
{\rm
First observe that the partial transpose of the $2\times N$
state
$\r=(\ket{00}+\ket{11})(\bra{00}+\bra{11})+\proj{0}\ox I_N$
has exactly one negative eigenvalue.
Next we consider the following family of $2\times N$ states
 \bea
 \r = \sum^{N-2}_{i=0} (\ket{0,i}+c_{i+1}\ket{1,i+1})(\bra{0,i}+c_{i+1}\bra{1,i+1}),
 \eea
where $0<c_1=\cdots=c_k<\cdots<c_{N-2}$, $1\le k<N-1$, and
$c_{N-1}=1$. Then $\r^\G = \sum^{N}_{i=1} M_i$ where
 \bea
 M_i &=& \proj{0,i+1}
 +c_{i}^2\proj{1,i}
 +c_{i+1} \ketbra{0,i+1}{1,i}
 +c_{i+1} \ketbra{1,i}{0,i+1},
 \quad i<N-2, \\
 M_{N-2} &=& \proj{0,1}
 +c_1\ketbra{0,1}{1,0}
 +c_1\ketbra{1,0}{0,1}, \\
 M_{N-1} &=& c_{N-2}^2\proj{1,N-2}
 +\ketbra{0,N-1}{1,N-2}
 +\ketbra{1,N-2}{0,N-1}, \\
 M_{N} &=& \proj{0,0}+\proj{1,N-1}
 \eea
are Hermitian matrices such that $M_i M_j = 0$ for $i\ne j$.
For $k\le i<N$ each $M_i$ has exactly one negative eigenvalue,
while for all other indexes $i$ the matrix $M_i\ge0$. Hence,
$\r^\G$ has exactly $N-k$ negative eigenvalues.
\hfill $\square$ }
 \eex


\acknowledgments

We thank Seung-Hyeok Kye for his comments on the first version
of this paper, and for supplying references \cite{hk04,ck12}.
The first author was mainly supported by MITACS and NSERC. The
CQT is funded by the Singapore MoE and the NRF as part of the
Research Centres of Excellence programme. The second author was
supported in part by an NSERC Discovery Grant.


\end{document}